\documentclass[aps,prd,groupedaddress,showpacs,nofootinbib,twocolumn
]{revtex4}
\usepackage{graphicx,amssymb,amsmath,bm}
\usepackage{subfigure}
\usepackage{multirow}

\newcommand{\be}{\begin{equation}}
\newcommand{\ee}{\end{equation}}
\newcommand{\bea}{\begin{eqnarray}}
\newcommand{\eea}{\end{eqnarray}}

\newcommand{\nn}{\nonumber}
\def\rf#1{(\ref{#1})}

\def\de#1/de#2{\frac{\partial {#1}}{\partial {#2}}}
\begin{document}

\title{Reconstructing exact scalar--tensor cosmologies via conformal transformations}

\author{  Stefano Vignolo$^{1}$\footnote{E-mail: vignolo@diptem.unige.it},\,  Sante Carloni$^{2}$\footnote{E-mail: sante.carloni@esa.int}\ and
Francesco Vietri$^{1}$\footnote{E-mail: vietri@diptem.unige.it}
}
\affiliation{ $^{1}$DIME Sez. Metodi e Modelli Matematici, Universit\`{a} di Genova\\
Piazzale Kennedy, Pad. D - 16129 Genova (Italia)\\$^{2}$ESA-Advanced Concept team, European Space Research Technology Center (ESTEC)\\
Keplerlaan 1, Postbus 299, 2200 AG Noordwijk The Netherlands.}

\begin{abstract}
We propose a new reconstruction method for scalar--tensor gravity based on the use of conformal transformations. The new method allows the derivation of a set of interesting exact cosmological solutions in brans Dicke gravity as well as other extensions of General Relativity.
\end{abstract}
\date{\today}
\maketitle

\tolerance=5000


\section{Introduction}
 In theoretical physics it  is common to have to deal with models which carry one or more indeterminate functions. Usually such unknown functions denote the lack of data on the physical phenomena related to aspect of the model under consideration. Faced with this kind of degeneration one has no other choice to investigate as many models as possible to try to select the ones which have the most useful features. 

This kind of problem is particularly felt in those sectors in which the experimental work is more difficult to perform like high energy physics and relativistic gravitation. An important example that belongs  to both these categories is related to the one of cosmologies with scalar fields and matter typical of the inflation scenario \cite{Inflation}. Since the key field  that drives the inflation, the inflaton,  has never been observed we have no information on the structure of its potential (i.e. its interactions)  and one is forced to deal ``in parallel'' with a great number of different models for the potential itself.

More recently, the same problem arose in the context of modifications of the gravitational interaction. Once the assumption of linearity and minimal coupling in the gravitational action has been abandoned, one is left with an infinite set of different theories whose phenomenology can be completely different. The problem is then to understand which of these models has interesting features and which has to be discarded.

An interesting approach to the resolution of this type of problem is given by the  so called reconstruction methods. First proposed by Lucchin and Matarrese in the 1980s \cite{Lucchin}   for the  determination of the free function(s) in a cosmological model, were further developed by Ellis and Madsen \cite{Ellis-Madsen} and by a number of other researchers using different techniques \cite{Reconstruction}.

The principle of reconstruction methods is simple, but very powerful: instead of trying to solve the cosmological equations for a set of forms of the unknown functions one finds the form of the unknown functions in the models that are compatible with a given  evolution of the relevant field of the theory.  

Reconstruction methods present a number of interesting advantages with respect to the traditional investigation approach. For example one can explore the full solution of a given theory without the necessity of introducing {\it ad hoc} assumptions which might hide interesting and unusual features of the theory like in the case of fourth order gravity \cite{Carloni:2010ph}. In addition, reconstruction methods allow the automatic selection of the model with the relevant/desirable characteristics without the need to explore one by one specific models. Finally the results obtained via reconstruction can successively used as a base for perturbative calculations which in turn can be a key for the comparison of a given theory of gravity with the available data.

In spite of these attractive features  the standard reconstruction methods have also  a series of limitations. They  are essentially due to the necessity of the resolution of high degree algebraic equations and/or high order differential equations. Such issues can seriously limit the effectiveness of the method itself.

In this paper we propose a new method, based on conformal transformations, able to ease this difficulty. The method is specifically designed for Brans-Dicke gravity, but we will see that with simple prescriptions it can be applied to other extensions of general relativity. It is well known that under conformal transformation one can map a given scalar tensor theory to Einstein gravity  with a minimally coupled scalar field \cite{CF-Papers}. Although the physical connection between conformal frames is not completely clarified, nothing forbids to use the conformal model as a working tool for reconstruction. In fact   in the case of Einstein gravity  with a minimally coupled scalar field  a very powerful reconstruction technique is known \cite{Ellis-Madsen}. The new method will exploit this technique to obtain a series of  exact solutions in the case of extended Brans-Dicke gravity and some more general types of scalar tensor theories. 

The paper is organized in the following way. In section II  we present the basic equations of an extended Brans Dicke theory and the properties of its conformal transformation. In section III we describe the reconstruction method for this specific theory. In section IV some examples are given. Section V describes the extension of this method to  a more general form of Brans-Dicke theory  and gives some additional examples. Section VI is dedicated to the conclusions.

Unless otherwise specified, natural units ($\hbar=c=k_{B}=8\pi G=1$) will be used throughout this paper, Latin indices run from 0 to 3. The symbol $\nabla$ represents the usual covariant derivative and $\partial$ corresponds to partial differentiation. We use the $+,-,-,-$ signature and the Riemann tensor is defined by
\begin{equation}
R^{a}{}_{bcd}=\Gamma^a{}_{bd,c}-\Gamma^a{}_{bc,d}+ \Gamma^e{}_{bd}\Gamma^a{}_{ce}-\Gamma^e{}_{bc}\Gamma^a{}_{de}\;,
\end{equation}
where the $\Gamma^a{}_{bd}$ are the Christoffel symbols (i.e. symmetric in the lower indices), defined by
\begin{equation}
\Gamma^a_{bd}=\frac{1}{2}g^{ae}
\left(g_{be,d}+g_{ed,b}-g_{bd,e}\right)\;.
\end{equation}
The Ricci tensor is obtained by contracting the {\em first} and the {\em third} indices
\begin{equation}\label{Ricci}
R_{cd}=g^{ab}R_{acbd}\;.
\end{equation}


\section{Brans-Dicke teories and conformal transformation}
Let us consider the action functional of a scalar-tensor theory
\begin{equation}\label{2.1}
{\cal A}\/(g,\varphi)=\int{\left[\sqrt{|g|}\left(\varphi{R}
-\frac{\omega_0}{\varphi}\varphi_i\varphi^i - U\/(\varphi)
\right)+ {\cal L}_m\right]\,ds},
\end{equation}
where $\varphi\/$ is the scalar field, $\varphi_i :=
\de\varphi/de{x^i}\/$ and $U\/(\varphi)\/$ is the potential of
$\varphi\/$. For $U\/(\varphi)=0\/$ such a theory reduces to the
standard Brans--Dicke theory \cite{brans}. The matter Lagrangian
${\cal L}_m\/$ is a function of the metric and some
matter fields $\psi\/$; $\omega_0\/$ is the so called Brans--Dicke
parameter. The field equations derived by varying with respect to
the metric and the scalar field are
\begin{eqnarray}\label{2.2}
\nn\varphi\/\left(R_{ij} -\frac{1}{2}R\/g_{ij}\right)&=&
\Sigma_{ij} + \frac{\omega_0}{\varphi}\left(
\varphi_i\varphi_j  - \frac{1}{2}\varphi_h\varphi^h\/g_{ij}
\right) \\&&+ \left({\nabla}_{j}\varphi_i - {\nabla}_h\varphi^h\/g_{ij} \right) -
\frac{U}{2}g_{ij},
\end{eqnarray}
and
\begin{equation}\label{2.3}
\frac{2\omega_0}{\varphi}{\nabla}_h\varphi^h + R -
\frac{\omega_0}{\varphi^2}\varphi_h\varphi^h - U' =0,
\end{equation}
where $\Sigma_{ij}:= -
\frac{1}{\sqrt{|g|}}\frac{\delta{\cal L}_m}{\delta g^{ij}}\/$ and
$U' :=\frac{dU}{d\varphi}\/$. Taking the trace of eq. \eqref{2.2} and using it to replace
$R\/$ in eq. \eqref{2.3}, one obtains the equation
\begin{equation}\label{2.4}
\left( 2\omega_0 + 3 \right)\/{\nabla}_h\varphi^h = \Sigma + \varphi U' -2U.
\end{equation}
Systems of equations \eqref{2.2}-\eqref{2.3} and \eqref{2.2}-\eqref{2.4} are therefore equivalent. Making use of well known properties of the Einstein and Ricci tensors, it is easily seen that eqs. \eqref{2.2} and \eqref{2.3} imply the standard conservation laws holding in General Relativity (GR) \cite{Koivisto}. Indeed, from the quadridivergence of \eqref{2.2} one gets
\be\label{2.5}
\nabla^j\Sigma_{ij} + \varphi_i\/\left( \frac{1}{2}R - \frac{1}{2}U' - \frac{\omega_0}{2\varphi^2}\varphi^j\varphi_j + \frac{\omega_0}{\varphi}\nabla^j\varphi_j\right) =0.
\ee
In view of \eqref{2.3}, eq. \eqref{2.5} implies 
\be\label{2.6}
\nabla^j\Sigma_{ij} =0.
\ee
On the contrary, supposing the conditions \eqref{2.6} and $\varphi_i \not = 0$ be true, eq. \eqref{2.5} amounts to eq. \eqref{2.3} which in such a circumstance is a consequence of eqs. \eqref{2.2}. 
\\
Now, let us perform the conformal transformation
\be\label{2.7}
\bar{g}_{ij} = \varphi\/g_{ij}.
\ee
The coefficients of the Levi--Civita connections $\Gamma$ and $\bar\Gamma$ associated with the metric tensors $g_{ij}$ and $\bar{g}_{ij}$ are related by the identity
\be\label{2.8}
\bar{\Gamma}_{ij}^{\;\;\;h}= {\Gamma}_{ij}^{\;\;\;h} +
\frac{1}{2\varphi}\de\varphi/de{x^j}\delta^h_i -
\frac{1}{2\varphi}\de\varphi/de{x^p}g^{ph}g_{ij} +
\frac{1}{2\varphi}\de\varphi/de{x^i}\delta^h_j.
\ee 
Making use of eq. \eqref{2.8}, it is a straightforward matter to verify that the Einstein tensors $G_{ij}:=\left(R_{ij} -\frac{1}{2}R\/g_{ij}\right)$ and $\bar{G}_{ij}:=\left(\bar{R}_{ij} -\frac{1}{2}\bar{R}\/\bar{g}_{ij}\right)$ induced by the metrics $g_{ij}$ and $\bar{g}_{ij}$ satisfy the relation
\bea\label{2.9}
\nn\bar{G}_{ij} &= &G_{ij} + \frac{3}{2\varphi^2}\left(
\varphi_i\varphi_j  - \frac{1}{2}\varphi_h\varphi^h\/g_{ij}\right)+\\
&&- \frac{1}{\varphi}\left({\nabla}_{j}\varphi_i - {\nabla}_h\varphi^h\/g_{ij} \right).
\eea
Therefore, passing from the Jordan frame $g_{ij}$ to the Einstein frame $\bar{g}_{ij}$, eqs. \eqref{2.2} can be rewritten in the equivalent form
\be\label{2.10}
\bar{G}_{ij} = \frac{1}{\varphi}\Sigma_{ij} + \frac{\Omega_0}{\varphi^2}\left(
\varphi_i\varphi_j  - \frac{1}{2}\varphi_h\varphi^h\/\bar{g}_{ij}
\right) - \frac{U}{2\varphi^2}\bar{g}_{ij},
\ee
where $\Omega_0 := \omega_0 + \frac{3}{2}$. In the same way, eq. \eqref{2.4} can be expressed in terms of the conformal metric $\bar{g}_{ij}$ as
\be\label{2.10bis}
\Omega_0\/\left(\varphi\bar{g}^{ij}\bar{\nabla}_i\varphi_j - \bar{g}^{ij}\varphi_i\varphi_j \right) =  \frac{1}{2}\left( \Sigma + \varphi U' -2U \right) .
\ee
Moreover, denoting by $T_{ij}$ the right--hand--side of \eqref{2.10} and taking eqs. \eqref{2.8} into account, a direct calculation shows the identity (the reader can find details of similar calculations in \cite{CV1,CV3})
\be\label{2.11}
\bar{\nabla}^j\/T_{ij} = \frac{1}{\varphi^2}\nabla^j\/\Sigma_{ij} - \frac{1}{\varphi^3}\varphi_i\/\left[\frac{\varphi}{2}U' -U + \frac{1}{2}\Sigma - \Omega_0\nabla_j\varphi^j \right],
\ee
which clarifies the relationship between the conservation laws in the Jordan and the Einstein frames. In particular, on the one hand we see that if eqs. \eqref{2.10} and \eqref{2.10bis} (or equivalently \eqref{2.2} and \eqref{2.4}) are satisfied then the condition $\bar{\nabla}^j\/T_{ij} =0$ amounts to $\nabla^j\/\Sigma_{ij} =0$; on the other hand, under the hypotheses $\nabla^j\/\Sigma_{ij} =0$ and $\varphi_i \not =0$, eqs. \eqref{2.10} imply eqs. \eqref{2.4} (or \eqref{2.10bis}).


\section{The reconstruction method}
In this section we show how to combine  the conformal transformation \eqref{2.7} and the procedure introduced in \cite{Ellis-Madsen}, to search for exact cosmological solutions of the theories \eqref{2.1}. The target of the method is the reconstruction of the potential function $U(\varphi)$ associated with the scale--factor of a given Friedmann-Lema\^{\i}tre-Robertson-Walker (FLRW) universe.

The procedure is based on the use of the Einstein--like equations only and the relationships between the conservation laws holding in the Jordan and Einstein frames play a crucial role. 

From now on, for simplicity we suppose that $\varphi >0$. In the opposite case $\varphi <0$, we could use the conformal factor $\psi = -\varphi$ instead of $\varphi$. Since eqs. \eqref{2.8} and \eqref{2.9} are quadratic in $\varphi$ and its derivatives, it is an easy matter to verify that the whole proposed procedure works equally well also in such a circumstance.

To start, let us consider the field equations \eqref{2.2} and \eqref{2.4} in presence of a cosmological perfect fluid with equation of state $p=\lambda\rho$ ($\lambda \in [0,1[$). The corresponding energy--momentum tensor is then given by
\be\label{3.1}
\Sigma_{ij} = (\rho + p)U_iU_j - p\,g_{ij},
\ee
with $U_iU_j\/g^{ij}=1$. We assume that the fluid satisfies the usual conservation laws
\be\label{3.2}
\nabla^j\/\Sigma_{ij} =0.
\ee
We search for FLRW cosmological models 
\begin{equation}\label{3.3}
d\bar{s}^2 = dt^2 - \frac{\bar{a}^{2}(t)}{\left(1 + kr^2/4\right)^2} \left(dr^2 + r^2\,d\theta^2 + r^2sin^{2}\theta\,d\varphi^2 \right),
\end{equation}
with $k=-1,0,1$, which are solutions of the conformally transformed Einstein--like equations \eqref{2.10}. 
We have to express in terms of the metric $\bar{g}_{ij}$ the conservation law \eqref{3.2}. In the Einstein frame the energy-impulse tensor $\Sigma_{ij}$ assumes the form 
\be\label{3.4}
\Sigma_{ij} = \frac{1}{\varphi}\left(\rho + p\right)\bar{U}_i\bar{U}_j - \frac{p}{\varphi}\bar{g}_{ij},
\ee
with $\bar{U}_i\bar{U}_j\bar{g}^{ij}=1$.
Making use of eq. \eqref{2.8}, it is easily seen that the conservation law $\nabla^i\Sigma_{ij} =0$ reduces to
\begin{equation}\label{3.5}
\frac{\dot\rho}{\rho} + (1+\lambda)\frac{\dot{\bar\tau}}{\bar\tau} - \frac{3}{2}\frac{\dot\varphi}{\varphi}(1+\lambda)=0,
\end{equation}
where $\bar{\tau}=:\bar{a}^3$. Setting $\Theta:= \varphi^{-\frac{3}{2}}$, one has $- \frac{3}{2}\frac{\dot\varphi}{\varphi}=\frac{\dot\Theta}{\Theta}$,
and then we can integrate eq. \eqref{3.5} as
\begin{equation}\label{3.6}
\rho =\rho_0\/(\bar{\tau}\Theta)^{-(1+\lambda)} \qquad \rho_0= {\rm const.}.
\end{equation}
Moreover, from the field equations \eqref{2.10} evaluated on \eqref{3.3} we get the two following equivalent equations
\begin{subequations}\label{3.7}
\be\label{3.7a}
3\bar{H}^2 + \frac{3k}{\bar{a}^2} = \frac{\rho}{\varphi^2} + \frac{\Omega_0}{2\varphi^2}\dot{\varphi}^2 - \frac{1},{2\varphi^2}U
\ee
\be\label{3.7b}
3\dot{\bar{H}} + 3\bar{H}^2 = -\frac{1}{2\varphi^2}\left(\rho + 3p\right) - \frac{\Omega_0}{\varphi^2}\dot{\varphi}^2 - \frac{1}{2\varphi^2}U,
\ee
\end{subequations}
The combination $2\eqref{3.7a} + \eqref{3.7b}$ yields
\begin{subequations}\label{3.8}
\be\label{3.8a}
\dot{\bar{H}} + 3\bar{H}^2 + \frac{2k}{\bar{a}^2} = - \frac{1}{2\varphi^2}U + \frac{2\rho}{3\varphi^2} - \frac{1}{6\varphi^2}\left(\rho + 3p\right),
\ee
while from the difference $\eqref{3.7a} - \eqref{3.7b}$ we obtain
\be\label{3.8b}
\frac{\dot{\varphi}^2}{\varphi^2} = \frac{2}{\Omega_0}\left(-\dot{\bar{H}} + \frac{k}{\bar{a}^2} - \frac{\rho}{3\varphi^2} - \frac{1}{6\varphi^2}\left(\rho + 3p\right)\right),
\ee
\end{subequations}
Eqs. \eqref{3.8} are, of course, equivalent to eqs. \eqref{3.7}. Now, choosing a desired scale factor $\bar{a}(t)$,
the  desired result is obtained via the following steps:
\begin{enumerate}
\item insert the solution \eqref{3.6} into \eqref{3.8b} and solve the resulting differential equation for $\varphi$;
\item the solution $\varphi(t)$ is inverted giving $t(\varphi)$;
\item inserting all the obtained solutions together with the relation $t(\varphi)$ into \eqref{3.8a}  we get the expression of the potential $U(\varphi)=U(t(\varphi))$.
\end{enumerate}
As it has been explained in section 2 (see eq. \eqref{2.11}), once the Einstein--like equations \eqref{2.10} and the conservation laws \eqref{3.2} are satisfied, the Klein--Gordon like equation \eqref{2.4} automatically holds, provided that $\dot\varphi \not =0$. The last step is expressing the Jordan frame $g_{ij}=\frac{1}{\varphi}\bar{g}_{ij}$ as a FLRW metric tensor; this can be made by performing the time trasformation 
\be\label{3.9}
d\bar{t}:=\frac{1}{\sqrt{\varphi}}dt.
\ee


\section{Examples}
\begin{figure}[htbp]
\begin{center}
\includegraphics[scale=0.65]{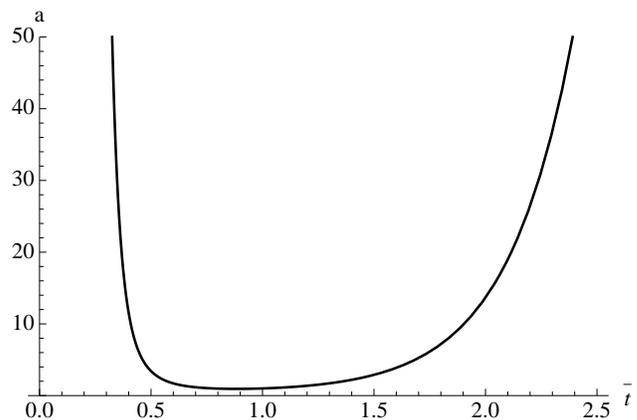}
\caption{An example of the behavior of the scale factor obtained in Example 2.}
\label{ex2}
\end{center}
\end{figure}

\noindent
{\bf Example 1.} In the Einstein frame and in presence of the scalar field $\varphi$ alone, let us consider a FLRW space--time \eqref{3.3} with constant scale factor $\bar{a}_0$. From eq. \eqref{3.8b} we derive the differential equation
\be\label{4.1}
\frac{\dot\varphi}{\varphi} = \pm A^2,
\ee  
where $A^2 := \frac{1}{\bar{a}_0}\sqrt{\frac{2k}{\Omega_0}}$ which imposes the condition $\frac{k}{\Omega_0}\geq 0$. A solution of \eqref{4.1} is
\be\label{4.2}
\varphi = B\exp\left(- A^2\/t\right),
\ee
with $B >0$. From \eqref{3.8a} we get the explicit expression for the associated potential $U(\varphi)$
\be\label{4.3}
U(\varphi)=-\frac{4k}{\bar{a}^2_0}\varphi^2.
\ee
In the Jordan frame the metric solution is then given by
\bea\label{4.4}
\nn ds^2 &=& B^{-1}\exp\left(A^2\/t\right)\,dt^2 - \frac{\bar{a}_0^2\/B^{-1}\exp\left(A^2\/t\right)}{{\left(1 + kr^2/4\right)^2}}\,\left(dr^2+ \right. \\ && \left.+ r^2\,d\theta^2 + r^2sin^{2}\theta\,d\varphi^2 \right).
\eea
Using the time-coordinate transformation
\be\label{4.5}
\bar{t} = \frac{2B^{-\frac{1}{2}}}{A^2}\exp\left(\frac{A^2}{2}\/t\right)+C,
\ee
the metric \eqref{4.4} assumes the FLRW form
\be\label{4.6}
ds^2 =d\bar{t}^2 - \frac{A^4\bar{a}_0^2(\bar{t}-C)^2}{4{\left(1 + kr^2/4\right)^2}}\,\left(dr^2 + r^2\,d\theta^2 + r^2sin^{2}\theta\,d\varphi^2 \right).
\ee
Therefore in the Jordan frame  the solution of the scale factor is
\be\label{4.7}
a(\bar{t})=\frac{A^2\bar{a}_0}{2}(\bar{t}-C),
\ee
while the scalar field
\be\label{4.7bis}
\varphi = \frac{4}{A^4}\left(\bar{t} -C\right)^{-2}.
\ee
This solution corresponds clearly to a Milne universe in which the scale factor increases  linearly in time. the classical Milne solution in GR, however, requires the open geometry of the space-like surfaces, while in this case the constant $k$ could have any sign as far as the ratio $\frac{k}{\Omega_0}$ remains positive\footnote{This prescription is necessary to avoid the scalar field to be null at finite time. In such situation the entire model and the conformal transformation is meaningless and we will neglect it.}. Solutions of this type are common in this type of theories and have been found independently with the dynamical system approach and are proven to be unstable \cite{Carloni:2007eu}.
\\
\\
{\bf Example 2.}  Still in the Einstein frame and in absence of fluid, we consider a spatially flat $(k=0)$ FLRW space--time undergoing an exponential expansion of the form $\bar{a}= \bar{a}_0\exp\left({\sigma t^2}\right)$, $\sigma >0$. Since one has $\bar{H}=\frac{\dot{\bar{a}}}{\bar a}=2\sigma\/t$ and $\dot{\bar{H}}=2\sigma$, eq. \eqref{3.8b} yields again the differential equation 
\be\label{4.8}
\frac{\dot\varphi}{\varphi} = \pm A^2,
\ee
where now $A^2 := \sqrt{-\frac{4\sigma}{\Omega_0}}$, being $\Omega_0 <0$. Eq. \eqref{4.8} still posseses solutions of the form \eqref{4.2}, from which we get the inverse relation 
\be\label{4.9}
t= -\frac{1}{A^2}\ln{\frac{\varphi}{B}}.
\ee
This last expression, inserted into eq. \eqref{3.8a},  yields the associated potential
\be\label{4.10}
U(\varphi)=-2\varphi^2\/\left[\frac{12\sigma^2}{A^4}\left(\ln\frac{\varphi}{B}\right)^2 + 2\sigma\right].
\ee
The metric in the Jordan frame is given by
\bea\label{4.11}
\nn ds^2 &=& B^{-1}\exp\left(A^2\/t\right)\,dt^2 - B^{-1}\exp\left(A^2\/t\right)\/\bar{a}^2_0 \times\\ &&\exp\left(2\sigma\/t^2\right)\,\left(dr^2 + r^2\,d\theta^2 + r^2sin^{2}\theta\,d\varphi^2 \right).
\eea
The time--coordinate change ($t\rightarrow\bar{t}$)
\be\label{4.12}
t= \frac{1}{A^2}\ln{\frac{BA^4}{4}\bar{t}^2},
\ee
allows to express the metric \eqref{4.11} as
\bea\label{4.13}
\nn ds^2 &=&d\bar{t}^2 - \frac{A^4\bar{a}^2_0}{4}\bar{t}^2\exp\left[\frac{2\sigma}{A^4}{\left(\ln{\frac{BA^4}{4}\bar{t}^2}\right)}^2\right]\,\left(dr^2+\right.\\&& \left. + r^2\,d\theta^2 + r^2sin^{2}\theta\,d\varphi^2 \right),
\eea
showing a scale factor with an accelerated behavior
\be\label{4.14}
a(\bar t)=\frac{A^2\bar{a}_0}{2}\bar{t}\exp\left[\frac{\sigma}{A^4}{\left(\ln{\frac{BA^4}{4}\bar{t}^2}\right)}^2\right].
\ee
The scalar field solution of the field equations becomes of the form
\be\label{4.14bis}
\varphi = \frac{4}{A^4}\bar{t}^{-2}.
\ee
The solution we obtained  is initially decreasing and, after reaching  a minimum, it grows again (See fig. \ref{ex2}). This kind of behavior can be associated to a bouncing cosmology. However this specific solution has the peculiarity that to have a very shallow minimum. Physically such type of bounce can be of interest because the permanence of the Universe in the high compressed state can have a non trivial effect on the relation between the physics of the contracting and expanding phase.

{\bf Example 3.} Let us consider again a spatially flat FLRW space--time with scale factor $\bar{a}=\bar{a}_0\/t^n$. We have $\bar{H}=\frac{\dot{\bar{a}}}{\bar a}=\frac{n}{t}$ and $\dot{\bar{H}}=-\frac{n}{t^2}$. If there is no coupled fluid, eq. \eqref{3.8b} becomes
\be\label{4.15}
\frac{\dot\varphi}{\varphi} =\pm\/A^2\frac{1}{t},
\ee
with $A^2 := \sqrt{\frac{2n}{\Omega_0}}$  and it admits the solution
\be\label{4.16}
\varphi= B\/t^{-A^2},
\ee
with $B >0$. The function \eqref{4.16} can be easily inverted, allowing to derive the corresponding potential  
\be\label{4.17}
U(\varphi)= - \frac{2n\left(3n-1\right)}{B^{\frac{2}{A^2}}}\varphi^{\frac{2A^2+2}{A^2}}.
\ee
The metric in the Jordan frame is expressed as
\bea\label{4.18}
\nn ds^2 &=& B^{-1}t^{A^2}\,dt^2 - B^{-1}\bar{a}^2_0\/t^{\left(A^2 + 2n\right)}\,\left(dr^2 +\right.\\&& \left. r^2\,d\theta^2 + r^2sin^{2}\theta\,d\varphi^2 \right).
\eea
Using the transformation of time--coordinate
\be\label{4.19}
\bar{t}= \frac{t^{\left(\frac{A^2}{2}+1\right)}}{B^{\frac{1}{2}}\left(\frac{A^2}{2}+1\right)},
\ee
and renaming some constants for simplicity, the metric \eqref{4.18} assumes the FLRW form
\be\label{4.20}
ds^2 =d\bar{t}^2 - {a}_0^2\/\bar{t}^{\frac{2\left(A^2+2n\right)}{A^2+2}}\,\left(dr^2 + r^2\,d\theta^2 + r^2sin^{2}\theta\,d\varphi^2 \right),
\ee
The corresponding scale factor
\be\label{4.21}
a(\bar{t})= a_0\/\bar{t}^{\frac{\left(A^2+2n\right)}{A^2+2}},
\ee
and  the scalar field \eqref{4.16} is 
\be\label{4.21bis}
\varphi = B\left[B^{\frac{1}{2}}\left(\frac{A^2}{2}+1\right)\bar{t}\right]^{-\frac{2A^2}{A^2+2}}.
\ee
This solution is a power law expansion whose exponent depends on the values of the parameter $n$ and $\Omega_0$. In particular it is evident that starting with a decreasing scale factor in the Einstein frame ($n<0$) one can only have  a contracting phase in the Jordan phase  for $\Omega _0<-1$  or a Friendmann-like expansion $\left(0<\frac{\left(A^2+2n\right)}{A^2+2}<1\right)$ if $-1<\Omega _0<0$. A Friendmann-like expansion  in the Einstein frame $0<n<1$ leads to a Friendmann-like expansion in the Jordan frame if $\Omega _0>0$ and an accelerated expansion in the Einstein frame leads to an accelerated expansion in the Jordan frame if $\Omega _0>0$\footnote{Also in this case as in Example 1 we consider only positive ratios between $\frac{n}{\Omega_0}$ to avoid the case of a null scalar field at finite time. }.


{\bf Example 4.} In this fourth example we consider a FLRW spatially flat space--time with scale factor $\bar{a}=\bar{a}_0\exp\left(\sigma\/t\right)$ in the presence of a dust fluid. In such a circumstance, from eq. \eqref{3.6} we have
\be\label{4.22}
\rho = \frac{\rho_0}{\bar{a}_0^3}\varphi^{\frac{3}{2}}\exp\left(-3\sigma\/t\right).
\ee
Inserting eq. \eqref{4.22} into \eqref{3.8b}, we obtain the differential equation
\be\label{4.23}
\varphi^{-\frac{3}{4}}\dot\varphi = - A^2\exp{\left(-\frac{3}{2}\sigma\/t\right)},
\ee
where $A^2 := \left(-\frac{\rho_0}{2\Omega_0\bar{a}_0^3}\right)^{\frac{1}{2}}$ with $\Omega_0 <0$. Setting $\frac{1}{B^2}:=\left(\frac{A^2}{6\sigma}\right)^4$, we can chose the solution
\be\label{4.24}
\varphi = \frac{1}{B^2}\exp\left(-6\sigma\/t\right).
\ee
The associated potential results to be
\be\label{4.25}
U(\varphi) = \left(\frac{B\rho_0}{\bar{a}_0^3}-6\sigma\right)\varphi^2,
\ee
while the metric in the Jordan frame is
\bea\label{4.26}
\nn ds^2 &=& B^2\exp\left(6\sigma\/t\right)\,dt^2 - B^2\exp\left(8\sigma\/t\right)\,\left(dr^2 +\right.\\&& \left. r^2\,d\theta^2 + r^2sin^{2}\theta\,d\varphi^2 \right).
\eea
The time transformation
\be\label{4.27}
\bar{t}= \frac{B}{3\sigma}\exp\left(3\sigma\/t\right),
\ee
allows to express the metric \eqref{4.26} in the final form
\be\label{4.28}
ds^2 =d\bar{t}^2 - {a}_0^2\bar{t}^{\frac{8}{3}}\,\left(dr^2 + r^2\,d\theta^2 + r^2sin^{2}\theta\,d\varphi^2 \right),
\ee
where $a_0$ is a suitable constant. The corresponding scale factor
\be\label{4.29}
a(\bar{t}) = a_0\/\bar{t}^{\frac{4}{3}},
\ee
undergoes an eccelerated expansion. The scalar field \eqref{4.24} assumes the form
\be\label{4.29bis}
\varphi = \frac{1}{9\sigma^2}\bar{t}^{-2}.
\ee
In this example, the solution found is a power law inflation in presence of dust fluid. Also this kind of solution has been found before and confirms again that non minimally coupled scalar tensor theories possess naturally solutions which can be used to model exotic types of inflation phases or dark energy.  
\\
\\
{\bf Example 5}
In the Einstein frame, we consider a spatially flat FLRW space--time with constant scalar factor $\bar{a}_0$, filled by a perfect fluid with equation of state $p=\frac{2}{3}\rho$. From eq. \eqref{3.6} we have
\be\label{4.30}
\frac{\rho}{\varphi^2}=\frac{\rho_0}{\bar{a}^5_0}\varphi^{\frac{1}{2}}.
\ee
From eq. \eqref{3.8b} we get the differential equation
\be\label{4.31}
\varphi^{-\frac{5}{4}}\dot\varphi = -A^2,
\ee
where $A^2= \left(\frac{5\rho_0}{3\Omega_0\bar{a}_0^5}\right)^{\frac{1}{2}}$ with $\Omega_0 <0$. A solution of \eqref{4.31} is
\be\label{4.32}
\varphi = \frac{1}{B^2}t^{-4},
\ee
where $\frac{1}{B^2}:=\left(\frac{A^2}{4}\right)^{-4}$. The associated potential is derived by eq. \eqref{3.8a} and it is 
\be\label{4.33}
U(\varphi)=\frac{\rho_0}{3\bar{a}_0^5}\varphi^{\frac{5}{2}}.
\ee
In the Jordan frame we have
\be\label{4.34}
ds^2 = B^2\/t^4\,dt^2 - \bar{a}_0^2\/B^2\/t^4\,\left(dr^2 + r^2\,d\theta^2 + r^2sin^{2}\theta\,d\varphi^2 \right).
\ee
Making use of the time transformation
\be\label{4.35}
\bar{t}= \frac{B}{3}t^3,
\ee
the metric \eqref{4.34} can be expressed in the FLRW form
\be\label{4.36}
ds^2 = d\bar{t}^2 - a_0^2\/\bar{t}^{\frac{4}{3}}\,\left(dr^2 + r^2\,d\theta^2 + r^2sin^{2}\theta\,d\varphi^2 \right),
\ee
with $a_0^2:=\bar{a}_0^2\/B^2\left(\frac{3}{B}\right)^{\frac{4}{3}}$. The corresponding scale factor is then 
\begin{equation}
a(\bar t)=a_0\/\bar{t}^{\frac{2}{3}},
\end{equation}
while the scalar field \eqref{4.32} becomes
\be\label{4.36bis}
\varphi = \frac{1}{B^2}\left(\frac{B}{3}\right)^{\frac{4}{3}}\bar{t}^{-\frac{4}{3}}.
\ee
This solution is exactly the one corresponding to the standard Friedmann solution in GR so that the \rf{2.1} and Einstein theory are indistinguishable at level of exact cosmology. The presence of such solution constitutes and additional confirmation of the mechanism for which in scalar tensor gravity can mimic General Relativity which has been found in \cite{GRLimit}. However this correspondence is true only at the level of the exact cosmology. The evolution of the linear perturbations on this background would show a different  behavior with respect to the standard  GR ones. This suggests the possibility to use the evolution of these perturbations to test these theories against the phenomena that are unexplained in the formation of linear structures.
\section{The case of generalized scalar tensor theories}
The considerations above were focused on an action close to the original Brans-Dicke model (which appeared originally in \cite{brans} without a potential). However it is well known that the original theory can be further generalized. The method proposed in the above section can be straightforwardly extended to these more general theories. In this section we will look to some examples of these more general cases.
 
Let us consider the action functional of  the scalar-tensor theory
\begin{equation}\label{ExtBD}
{\cal A}\/(g,\varphi)=\int{\left[\sqrt{|g|}\left(\varphi{R}-\frac{\omega(\varphi)}{\varphi}\varphi_i\varphi^i - U\/(\varphi)
\right)+ {\cal L}_m\right]\,ds},
\end{equation}
the associated  field equations are
\begin{eqnarray}\label{JF}
\nn \varphi\/\left(R_{ij} -\frac{1}{2}R\/g_{ij}\right)&=&
\Sigma_{ij} + \frac{\omega (\varphi)}{\varphi}\left(
\varphi_i\varphi_j  - \frac{1}{2}\varphi_h\varphi^h\/g_{ij}
\right) +\\&&+ \left({\nabla}_{j}\varphi_i - {\nabla}_h\varphi^h\/g_{ij} \right) -
\frac{U}{2}g_{ij},
\end{eqnarray}
and the Klein Gordon equation is
\begin{equation}
\left( 2\omega(\varphi) + 3 \right)\/{\nabla}_h\varphi^h+ \omega'(\varphi)\varphi_i\varphi^i= \Sigma +
\varphi U' -2U= T_{ij}.
\end{equation}
Under the conformal transformation (\ref{2.7}) these equations reduce to the Einstein equations minimally coupled to a scalar field:
\begin{equation}\label{EFEq}
\bar{R}_{ij} -\frac{1}{2}\bar{R}\/g_{ij}=
\bar{\Sigma}_{ij}(\phi(\varphi),\bar{g}) + 
\phi_i\phi_j  - \frac{1}{2}\phi_h\phi^h\/g_{ij} -
\frac{W(\phi)}{2}\bar{g}_{ij},
\end{equation}
where
\begin{equation}\label{FieldRedefBD}
\phi=\int\sqrt{\frac{3-2\,\omega(\varphi)}{\varphi^2}} d\varphi \qquad W(\phi)=\left.\frac{U(\varphi)}{\varphi}\right|_{\varphi=\varphi(\phi)}.
\end{equation}
Like in the case of constant $\omega(\varphi)$ it is easy to show that the condition $\bar{\nabla}^j\/T_{ij} =0$ amounts to $\nabla^j\/\Sigma_{ij} =0$.
The cosmological equations associated to (\ref{EFEq}) are
\begin{subequations}\label{FieldRedef}
\be\label{FieldRedefa}
3\bar{H}^2 + \frac{3k}{\bar{a}^2} = \frac{\rho}{\phi^2} + \frac{1}{2}\dot{\phi}^2 -W,
\ee
\be\label{FieldRedefb}
3\dot{\bar{H}} + 3\bar{H}^2 = -\frac{1}{2\phi^2}\left(\rho + 3p\right) - \frac{1}{2}\dot{\phi}^2 -W,
\ee
\end{subequations}
At this point one can use the method of \cite{Ellis-Madsen} to reconstruct a solution for the above equations and use the \rf{FieldRedef} to obtain the solution for the field and the potential.

\subsection{Two simple examples.}\label{Ex_ScTn}

Let us consider two choices  of $\omega(\varphi)$ for which the integral in \rf{FieldRedefBD} is straighforward:
\begin{equation}
\omega_1(\varphi)=\frac{3-\varphi^{2(\alpha+1)}}{2},\qquad \omega_2(\varphi)=\frac{3-\varphi^{2}e^{-2\alpha\varphi}}{2},
\end{equation}
which, setting the integration constant to zero correspond to
\begin{equation}
\phi=\frac{\varphi^{\alpha+1}}{\alpha+1} \qquad \phi=\frac{e^{-\alpha \varphi}}{\alpha},
\end{equation}
respectively.

Considering  for example the solution 
\begin{equation}
\bar{a}(t)= \bar{a}_0 e^{ \beta t},\quad \phi= \frac{2 k}{\bar{a}_0^2\,\beta}e^{-\beta t},\quad U(\phi)=\beta^{2} \left[3 +\phi^2\right],
\end{equation}
given in \cite{Ellis-Madsen}, we obtain for $\omega_1$ 
\begin{equation}
a(\bar{t})= \frac{2k (\alpha +1)^{\alpha +1} }{(-1)^{\alpha }\bar{a}_0\, \beta ^{\alpha +1}  }\ \bar{t}^{-\alpha },\qquad\varphi=-\frac{\beta  }{\alpha +1}\,\bar{t}\label{SolConf1},
\end{equation}
and 
\begin{equation}
U(\varphi)=\beta ^2 \left(3+\frac{\varphi^{2 \alpha +2}}{(\alpha +1)^2}\right).\label{VConf1}
\end{equation}
The solution obtained represents therefore a power law expansion for the cosmological model. Note that, depending on the choice of the definition of the field, one is able to modify the behavior of the cosmological model.

\begin{figure}[htbp]
\begin{center}
\includegraphics[scale=0.65]{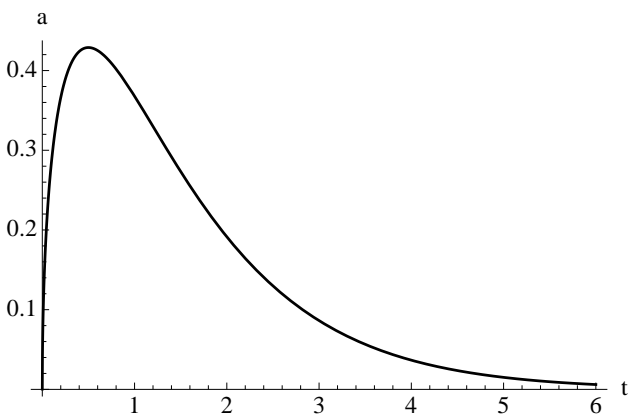}
\caption{An example of the behavior of the scale factor obtained with the function $\omega_2$ in section \ref{Ex_ScTn}.}
\label{ex2_2}
\end{center}
\end{figure}

For $\omega_2$ we have
\begin{equation}
a(\bar{t})= \frac{2 k}{\bar{a}_0}\sqrt{\frac{ 2 \, \alpha}{ \beta }}\, \bar{t}^{\,1/2} e^{-\sqrt{2 \alpha\, \beta }\,\bar{t}^{\,1/2}},\qquad\varphi=\sqrt{\frac{ 2\, \beta}{\alpha}}\, \bar{t}^{\,1/2},\label{SolConf2}
\end{equation}
and 
\begin{equation}
U(\varphi)=\beta ^2 \left(3+\frac{e^{-2 \alpha  \varphi }}{\alpha ^2}\right)\label{VConf2},
\end{equation}
which represent a universe which expands up to a maximum size and than recollapses asymptoticallly (see figure \ref{ex2_2}).

Note that there is a degeneration in the formulas above in the sense that  a solution for $\phi$ maps in different solutions for $\varphi$ and $U$ depending on the choice of the form of $\omega(\varphi)$.  Therefore for each solution in the Einstein frame there is a number of pairs theory--solution in the Jordan frame. This makes the conformal reconstruction method more powerful than the standard one which usually returns only one pair theory--solution.
\section{Conclusions}
In this paper we have proposed a new reconstruction technique for scalar tensor gravity based on conformal transformations. Using the reconstruction technique of \cite{Ellis-Madsen} in the (conformal) Einstein frame and mapping the reconstructed solution to the (original) Jordan frame, one is able to obtain a series of new exact solutions for different kinds of non minimally coupled scalar tensor theories.  

The necessity of the definition of such a method appears clearly when one considers the complexity of  eqs. \eqref{2.2}. Even in the cosmological framework these equations are very difficult to solve analytically. For example, the counterpart of \rf{3.8b}  in the Jordan frame is
\be\label{3.10}
\frac{1}{2}\frac{\ddot\varphi}{\varphi} + \frac{\omega_0}{2}\frac{{\dot\varphi}^2}{\varphi^2} - \frac{1}{2}\frac{\dot a}{a}\frac{\dot\varphi}{\varphi} + \frac{\rho}{3\varphi} + \frac{1}{6\varphi}\left(\rho + 3p\right) + \dot{H} - \frac{k}{a^2} =0.
\ee
which is much more difficult to integrate and limits strongly the power of a direct reconstruction method.

In our approach the physical meaning of the solution in the Einstein frame is irrelevant as the conformal frame only plays the role of auxiliary problem. The advantage of this approach lays in the fact that in the conformal frame the equations related to the reconstruction are much easier to be solved.

Among the examples discussed in the case of Brans-Dicke theory with a non vanishing potential we found an interesting cosmology representing a ``loitering bounce'' i.e. a  bounce with a very flat minimum. During this phase the universe has a relatively short size and it would be interesting to investigate further the type of processes that might take place during such phase  and the specific signatures that they would leave on the late time evolution of the cosmological model.  Another interesting solution for the Brans-Dicke theory is the one given in example 5 in which a cosmology presenting a mix of dust and radiation evolves according with the classical Friedmann expansion law.  This mimicking phenomenon is known since some years and our techniques allows a further confirmation of it in the case of Brans-Dicke gravity. 

The technique presented is independent of the choice of the non minimal coupling of the scalar field and can be therefore applied to Brans-Dicke theories with generalized kinetic terms.  In this framework it becomes immediately apparent that  given a single solution in the Einstein frame one can deduce a number of pairs theory-solution. This makes the method even more interesting in terms of a solution generator and reveals at the same time the consequence of the lack of constraints on the scalar field  sector of these theories. A change in the definition of such scalar can lead to a radical change in the expansion law. 

Although the method proposed has been specifically designed for the Brans-Dicke theory with a potential, it is easy to see that the same algorithm can be used to obtain exact solutions for other interesting non standard theories of gravitation. For example a Brans-Dicke theory with generalized kinetic term can be mapped to all the other formulations of scalar--tensor gravity via a redefinition of the scalar field.  This mean that for every reconstructed solution in GR one can obtain a pair theory solution for every type of scalar tensor action.

Another interesting example concerns $f(R)$-gravity. This class of theories can be mapped to a Brans-Dicke-like theory with no kinetic term and a nontrivial potential (O' Hanlon Lagrangian) \cite{Faraoni:1998qx} via the transformation
\begin{equation} \label{SF-Trans}
\varphi=f'(R) \qquad \mbox{and} \qquad U(\varphi)=R(\varphi)\varphi-f(R(\varphi)),
\end{equation}
and one could therefore find the form of the function $f$ and the solution that corresponds to a given solution in the Einstein frame. This connection, however, entails  additional difficulties due to the fact that one has to be able to  invert the relation $R=R(t)$ deduced for the scale factor solution and that the second  of \rf{SF-Transf} acts as a constraint. In principle therefore not all the Einstein frame solutions can be mapped to solutions of $f(R)$-gravity and this shows a difference in the set of solutions of the two theories.

The same hold for  $f(R)$-gravity  with torsion. These theories maps to Brans-Dicke theory with potential and $\omega(\varphi)=-3/2$. This value for $\omega$ is pathologic because it eliminates the D'Alembertian in \rf{2.4} but such pathology is irrelevant for our method. In this case one should reconstruct  also the torsion tensor, but this can be found  via the relation
\begin{equation}
\label{2.1b}
T_{ab}^{\;\;\;c}
=\frac{1}{2}\frac{ \partial_p \varphi}{\varphi}
\left(\delta^{p}_{b}\delta^{c}_{a}-\delta^{p}_{a}\delta^{c}_{b}\right)
\end{equation}
that connects the torsion tensor to the scalar field $\varphi$.

We can conclude, therefore, that since all these theories can be mapped to a Brans-Dicke-like theory  one can reconstruct (minding additional constraints) pairs theory-solution starting form the ones of the conformal frame. This fact gives to the simple method described in this work a much ample range of application than it was initially expected. We believe that this method will be able to provide a set of exact cosmological solutions which will improve our understanding of the cosmology of these large class of theories.


\end{document}